\documentclass[runningheads]{llncs}
\usepackage[T1]{fontenc}
\usepackage{graphicx}
\usepackage{amsmath}
\usepackage{amssymb}
\usepackage{booktabs}
\usepackage{url}
\usepackage{multirow}
\usepackage{xcolor,colortbl}
\usepackage{hyperref}
\usepackage{orcidlink}

\begin{document}
\title{Cross-Sensory Brain Passage Retrieval: Scaling Beyond Visual to Audio}

\author{Niall McGuire\inst{1}\,\orcidlink{0009-0005-9738-047X} \and
Yashar Moshfeghi\inst{1}\,\orcidlink{0000-0003-4186-1088}}

\authorrunning{N. McGuire and Y. Moshfeghi}

\institute{University of Strathclyde, NeuraSearch Laboratory, Glasgow, UK\\
\email{\{niall.mcguire,yashar.moshfeghi\}@strath.ac.uk}}

\maketitle              % typeset the header of the contribution

\begin{abstract}
Query formulation from internal information needs remains fundamentally challenging across all Information Retrieval paradigms due to cognitive complexity and physical impairments. Brain Passage Retrieval (BPR) addresses this by directly mapping EEG signals to passage representations without intermediate text translation. However, existing BPR research exclusively uses visual stimuli, leaving critical questions unanswered: Can auditory EEG enable effective retrieval for voice-based interfaces and visually impaired users? Can training on combined EEG datasets from different sensory modalities improve performance despite severe data scarcity? We present the first systematic investigation of auditory EEG for BPR and evaluate cross-sensory training benefits. Using dual encoder architectures with four pooling strategies (CLS, mean, max, multi-vector), we conduct controlled experiments comparing auditory-only, visual-only, and combined training on the Alice (auditory) and Nieuwland (visual) datasets. Results demonstrate that auditory EEG consistently outperforms visual EEG, and cross-sensory training with CLS pooling achieves substantial improvements over individual training: 31\% in MRR (0.474), 43\% in Hit@1 (0.314), and 28\% in Hit@10 (0.858). Critically, combined auditory EEG models surpass BM25 text baselines (MRR: 0.474 vs 0.428), establishing neural queries as competitive with traditional retrieval whilst enabling accessible interfaces. These findings validate auditory neural interfaces for IR tasks and demonstrate that cross-sensory training addresses data scarcity whilst outperforming single-modality approaches\footnote{Code: \url{https://github.com/NiallMcguire/Audio_BPR}}.

\end{abstract}

\section{Introduction}

The translation of internal information needs into explicit textual queries represents one of the most enduring challenges in Information Retrieval (IR), where users must externalise often uncertain or ill-defined cognitive states through conventional interaction mechanisms~\cite{belkin1980anomalous,ingwersen1996cognitive,moshfeghi2016understanding}. This cognitive burden persists across all IR paradigms, from traditional ad hoc retrieval~\cite{robertson2009probabilistic} to modern chatbots and conversational search powered by large language models~\cite{radlinski2017theoretical,lupart2025disco}. The complexity of information needs realisation manifests through multiple barriers: cognitive challenges where users struggle to articulate vague or emerging information needs~\cite{kuhlthau2005information,taylor1968question}, linguistic barriers in translating thoughts into effective query terms~\cite{furnas1987vocabulary}, and physical or mental impairments that fundamentally limit users' ability to interact with traditional input mechanisms~\cite{wolpaw2002brain}. Brain Passage Retrieval (BPR) was recently proposed to address these challenges~\cite{mcguire2025towards}, rather than attempting to decode brain signals into intermediate text queries, BPR directly maps EEG signals recorded during reading to dense passage representations in a shared semantic space.

However, existing BPR research focuses exclusively on visual stimuli recorded during reading tasks~\cite{hollenstein2018zuco,hollenstein2019zuco}, creating critical limitations that constrain both its applicability and ecological validity. First, no evidence exists for whether auditory EEG can serve as effective query representations, despite the growing importance of audio-based retrieval scenarios. Conversational search systems~\cite{radlinski2017theoretical}, voice-activated interfaces, and retrieval over spoken content (podcasts, audiobooks) represent contexts where visual stimuli are absent, yet users still form information needs. Understanding whether auditory neural patterns can support retrieval is essential for Brain-Machine Interfaces (BMIs) that operate across diverse interaction modalities. Second, whether training on diverse EEG datasets from different sensory modalities improves retrieval performance remains unexplored, which would be valuable given the severe scarcity of EEG training data. This capability would be valuable given the severe scarcity of EEG training data, where collection requires specialised equipment, controlled conditions, and extensive participant time~\cite{hollenstein2018zuco,bhattasali2020alice,roy2019deep}.

To address these gaps, we present the first systematic investigation of auditory EEG for BPR and evaluate the potential of cross-sensory training to enhance retrieval performance. Employing a dual encoder architecture~\cite{karpukhin2020dense} with EEG query processing components, we implement and compare four semantic aggregation strategies: CLS token pooling~\cite{devlin2018bert}, mean pooling~\cite{lin2020distilling}, max pooling~\cite{lin2023aggretriever}, and multi-vector representations~\cite{khattab2020colbert}. Using the Alice dataset~\cite{bhattasali2020alice} for auditory stimuli and the Nieuwland dataset~\cite{nieuwland2019distinguishing} for visual stimuli, we conduct controlled experiments comparing individual auditory-only, individual visual-only, and combined cross-sensory training. This design enables direct assessment of whether cross-sensory learning enables retrieval across different sensory contexts whilst addressing limited EEG training resources, advancing toward more versatile brain-machine interfaces for information retrieval~\cite{moshfeghi2025brain}.

\begin{figure}
    \centering
    \includegraphics[width=\linewidth]{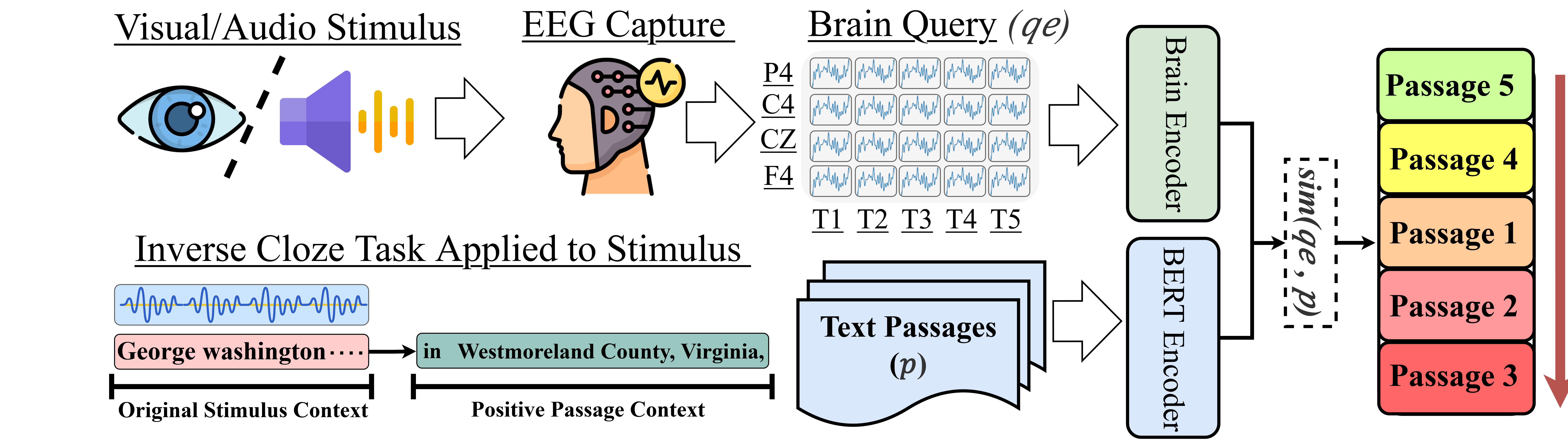}
    \caption{Brain Passage Retrieval framework: EEG signals recorded during visual or auditory stimulus presentation serve as brain queries ($\mathbf{q}_e$), which are encoded by a dual-encoder architecture alongside text passages ($p$). The encoded representations are mapped to a shared semantic space where similarity computation produces relevance scores for passage ranking and retrieval.}
    \label{fig:bpr_overview}
\end{figure}

\section{Related Work}

The intersection of neuroscience and information retrieval has evolved from foundational studies of neural IR processes towards sophisticated brain-machine interfaces capable of direct semantic decoding. Early neuroimaging investigations established that core IR concepts manifest as distinct brain signatures~\cite{moshfeghi2016understanding,gwizdka2019introduction}, with Moshfeghi et al.~\cite{moshfeghi2016understanding,moshfeghi2013understanding} demonstrating that information need formation produces specific activation patterns in the posterior cingulate cortex and developing neuropsychological models of information need realization. Subsequent investigations have further characterised the neural signatures of information need realization~\cite{paisalnan2021towards,moshfeghi2019towards} and satisfaction during search~\cite{paisalnan2022neural}. Building on this foundation, subsequent research has characterised the metacognitive states preceding search behaviour~\cite{michalkova2022information,michalkova2024understanding} and revealed consistent neurological correlates of relevance across users through EEG studies~\cite{allegretti2015relevance,pinkosova2020cortical,pinkosova2022revisiting}. Further investigations have examined the neural correlates of satisfaction during information seeking~\cite{paisalnan2021towards,paisalnan2022neural} and the role of neural correlates in query term specificity~\cite{kangassalo2019users}. Eugster et al. demonstrated term-relevance prediction directly from EEG signals with 17\% improvements over baseline. Ji et al.~\cite{ji2024characterizing} conducted a multimodal physiological study across four search stages, finding that cognitive load and affective arousal vary significantly throughout the process with the highest responses during relevance judgement. Complementary work on mental workload classification from EEG signals~\cite{mcguire2023ensemble} has demonstrated the effectiveness of deep learning and ensemble approaches for decoding cognitive states from neural signals. More recently, McGuire and Moshfeghi~\cite{mcguire2024prediction} achieved 90\% accuracy in predicting information need realisation from EEG signals, demonstrating the practical potential of neural decoding for IR applications.

Translating these neuroscience findings into functional brain-machine interfaces for IR has progressed through distinct technological approaches. Early implementations utilised Steady-State Visually Evoked Potentials for character selection~\cite{chen2022web}, whilst Eugster et al.~\cite{eugster2016natural} demonstrated direct document recommendation from brain signals during naturalistic reading. Kauppi et al.~\cite{kauppi2015towards} advanced this direction by decoding image relevance from MEG signals for brain-activity-controlled retrieval. To enable more expressive communication, Wang and Ji~\cite{wang2022open} introduced EEG-to-Text (EEG2Text), formulating brain signal decoding as a machine translation problem. However, this approach faces fundamental limitations: Jo et al.~\cite{jo2024eeg} demonstrated that current EEG-to-text models rely on teacher forcing and exhibit similar performance with noise inputs, suggesting memorisation rather than meaningful semantic mapping. Research on semantic alignment between brain signals and transformer models~\cite{lamprou2022role} has investigated how linguistic features influence the mapping between neural and computational representations, revealing the complexity of establishing robust brain-to-text correspondence.

Addressing these limitations, McGuire and Moshfeghi~\cite{mcguire2025towards} proposed Brain Passage Retrieval (BPR), which bypasses intermediate text generation by leveraging dense retrieval architectures~\cite{karpukhin2020dense,khattab2020colbert} and contrastive learning to project EEG signals directly into passage representation spaces. While Jo et al.'s findings~\cite{jo2024eeg} suggest that explicit text reconstruction from EEG is unreliable, direct mapping to passage embeddings circumvents this bottleneck by learning semantic alignment without requiring intermediate linguistic decoding, thus avoiding the teacher forcing and memorisation issues that plague translation-based approaches. Complementary approaches have explored using brain signals for query augmentation~\cite{ye2024query} and language generation from brain recordings~\cite{ye2024language,ye2024generative}, demonstrating alternative pathways for leveraging neural signals in IR tasks. Zhang et al.~\cite{zhang2024improving} further demonstrated the application of brain signals to specialized retrieval tasks such as legal case retrieval. Whilst promising, existing BPR research has been constrained to visual stimulus presentation during reading tasks~\cite{hollenstein2018zuco,hollenstein2019zuco}, leaving auditory EEG entirely unexplored despite cognitive neuroscience research demonstrating that auditory and visual processing engage distinct neural pathways with different semantic representation patterns~\cite{hickok2007cortical,huth2016natural}. Furthermore, whether combining training data across different EEG collection paradigms improves model performance remains uninvestigated, which is particularly important given data scarcity challenges ~\cite{huth2016natural}. This gap represents both a theoretical limitation in understanding cross-sensory neural information processing and a practical barrier to developing inclusive brain-machine interfaces for voice-based interactions and users with visual impairments.

\section{Methodology}
\label{methodology}

\subsection{Task Formulation}
\label{task_formulation}

We investigate the comparative effectiveness of EEG signals captured during auditory versus visual stimulus presentation within the Brain Passage Retrieval (BPR) framework~\cite{mcguire2025towards}. This framework eliminates intermediate text translation by directly mapping EEG signals to dense passage representations in a shared semantic space, avoiding the limitations identified in EEG-to-text approaches~\cite{jo2024eeg}.

Let $\mathcal{D} = \{w_1, w_2, \ldots, w_L\}$ denote a passage of length $L$ words. During stimulus presentation, EEG signals are recorded as $\mathbf{X} = [\mathbf{x}_1, \mathbf{x}_2, \ldots, \mathbf{x}_L] \in \mathbb{R}^{L \times D}$, where $\mathbf{x}_i \in \mathbb{R}^D$ represents the neural response to word $w_i$ across $D = C \times T$ features from $C$ channels and $T$ temporal samples. Given a corpus $\mathcal{C} = \{p_1, p_2, \ldots, p_N\}$ of $N$ passages, the objective is to retrieve the top-$k$ most relevant passages for an EEG query $\mathbf{q} \subseteq \mathbf{X}$.

The approach employs a dual-encoder architecture:
\begin{align}
f_{\text{eeg}}&: \mathbb{R}^{L \times D} \rightarrow \mathbb{R}^d \\
f_{\text{text}}&: \mathcal{V}^* \rightarrow \mathbb{R}^d
\end{align}
where $f_{\text{eeg}}$ encodes EEG sequences, $f_{\text{text}}$ encodes text passages, and both map to a shared $d$-dimensional embedding space. Retrieval scores are computed as:
\begin{equation}
\text{sim}(\mathbf{q}, p) = f_{\text{eeg}}(\mathbf{q})^\top f_{\text{text}}(p)
\end{equation}

\begin{table}[t]
\centering
\caption{Dataset statistics and cross-sensory comparison for balanced EEG corpora. Visual modality uses Nieuwland dataset, audio modality uses Alice dataset. Lexical overlap computed as Jaccard similarity.}
\label{tab:dataset_stats}
\footnotesize
\begin{tabular}{lcc|cc}
\toprule
& \multicolumn{2}{c}{\textbf{Training}} & \multicolumn{2}{c}{\textbf{Validation}} \\
\textbf{Metric} & \textbf{Visual} & \textbf{Audio} & \textbf{Visual} & \textbf{Audio} \\
\midrule
Total queries & 1,200 & 1,200 & 300 & 300 \\
Total words & 28,989 & 32,469 & 7,312 & 8,415 \\
Unique words & 674 & 543 & 627 & 519 \\
Avg. query length & 6.6 & 7.5 & 6.7 & 7.8 \\
Avg. passage length & 17.5 & 19.6 & 17.7 & 20.2 \\
\midrule
\multicolumn{5}{l}{\textbf{cross-sensory Overlap}} \\
\midrule
Lexical similarity & \multicolumn{2}{c}{0.175} & \multicolumn{2}{c}{0.184} \\
\bottomrule
\end{tabular}
\end{table}

\subsection{Datasets and Materials}
\label{datasets_materials}

Training neural retrieval models requires substantial quantities of EEG-text paired data with corresponding query-document relationships~\cite{karpukhin2020dense}. However, no existing EEG dataset provides both auditory and visual stimulus presentation modalities with sufficient passage content for retrieval experiments. Moreover, unlike traditional IR datasets~\cite{nguyen2016ms}, no EEG datasets contain predefined query-document pairs necessary for retrieval training. 

To address these limitations, we utilise two complementary datasets that provide the requisite modality coverage whilst maintaining comparable experimental properties. For auditory stimulus presentation, we employ the Alice EEG dataset~\cite{bhattasali2020alice}, which contains EEG recordings from participants listening to the first chapter of \textit{Alice's Adventures in Wonderland}. For visual stimulus presentation, we utilise the Nieuwland dataset~\cite{nieuwland2019distinguishing}, which provides EEG recordings from participants reading narrative passages presented word-by-word at controlled presentation rates. The visual stimuli consisted of contextually rich discourse passages designed to elicit natural language comprehension processes, with words presented centrally on screen under standardised viewing conditions to ensure consistent visual processing across participants.

These datasets were selected for their comparable experimental properties essential for cross-sensory comparison: similar vocabulary sizes (Alice: 2,129 unique tokens; Nieuwland: 2,247 unique tokens), balanced participant populations (Alice: 49 subjects; Nieuwland: 51 subjects), and comparable data collection protocols with naturalistic language comprehension paradigms. The ZuCo dataset~\cite{hollenstein2018zuco,hollenstein2019zuco}, utilised in previous BPR investigations~\cite{mcguire2025towards}, lacks these balanced properties and provides only visual EEG recordings. Both selected datasets employed standardised EEG preprocessing pipelines, ensuring data quality suitable for cross-sensory neural analysis. Critically, while these datasets employ different source texts (Alice vs. Nieuwland narratives), this creates a challenging test: if combined training improves performance despite different texts and modalities, this would suggest that increasing dataset diversity (even across different collection paradigms) can benefit BPR models. The low lexical overlap between datasets (Jaccard similarity: 0.175-0.184, \ref{tab:dataset_stats}) creates a challenging scenario: if combined training improves both modalities despite minimal shared vocabulary, this would potentially suggest that models learn abstract semantic encodings from diverse neural patterns rather than exploiting text-specific lexical features.

To generate query-document pairs from these naturalistic reading datasets, we adapt the inverse cloze task (ICT) framework~\cite{lee2019latent,chang2020pre} following the implementation detailed in~\cite{mcguire2025towards}. ICT operates by treating text spans as implicit queries whilst considering their surrounding context as relevant passages. Specifically, given a passage $\mathcal{D}$ of length $L$ words with corresponding EEG signals $\mathbf{X}$, we: (1) randomly select a starting position and extract a contiguous span comprising 30\% of the passage as the query $q_e$ along with its EEG signals; (2) with probability $p_{\text{mask}} = 0.9$, remove this query span from the original passage to create the positive training passage, or with probability $1-p_{\text{mask}}$, retain the span in the passage. This probabilistic masking strategy encourages the model to learn robust semantic representations rather than relying on exact lexical overlap between queries and passages—when the query span is retained, the model cannot simply match surface forms but must learn deeper semantic relationships. This follows established practices in dense retrieval for preventing lexical shortcut learning~\cite{karpukhin2020dense}. See Table \ref{tab:dataset_stats} for dataset overview.

\begin{figure}[h]
    \centering
    \includegraphics[width=0.7\linewidth]{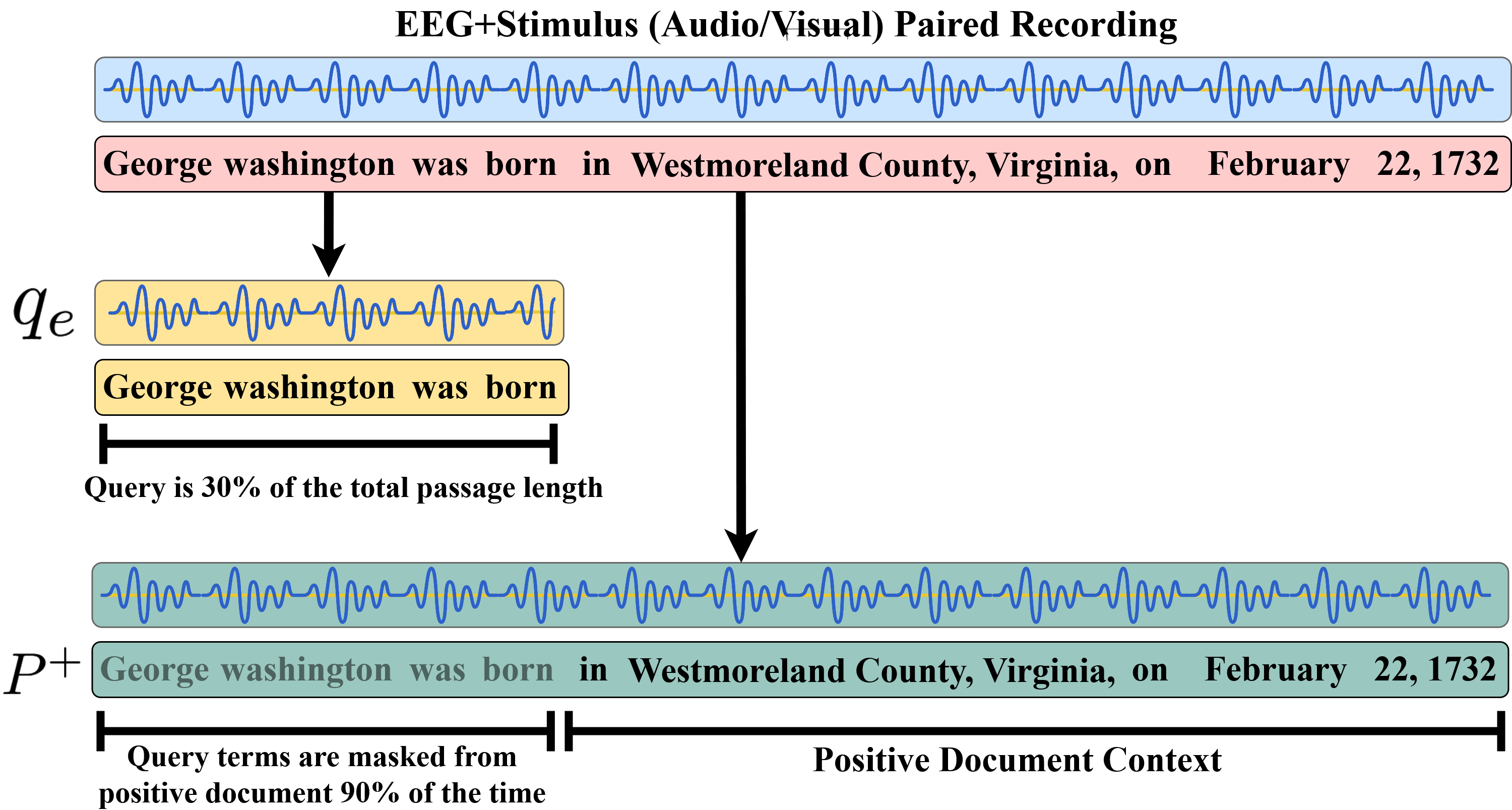}
    \caption{Inverse Cloze Task (ICT) overview: During stimulus presentation (audio or visual), EEG signals are recorded synchronously with text comprehension. Query spans $q_e$ comprising 30\% of the total passage length are extracted along with their corresponding neural responses. Positive passages $P^+$ are created by masking the query span from the original passage with probability $p_{\text{mask}} = 0.9$.}
    \label{fig:placeholder}
\end{figure}

\subsection{Model Architecture}
\label{model_architecture}

Our model implements a dual-encoder architecture adapted from dense retrieval frameworks~\cite{karpukhin2020dense} with specialised components for EEG processing. The text encoder employs BERT-base-uncased~\cite{devlin2018bert} as the base model, chosen for its proven effectiveness in dense retrieval tasks whilst maintaining robust semantic representation capabilities. We freeze all pre-trained BERT parameters throughout training to focus adaptation efforts exclusively on the EEG modality, ensuring stable text representations whilst enabling effective cross-sensory alignment. Only the EEG encoder parameters are trained from scratch to learn the cross-modal mapping between neural signals and the frozen text embedding space. The EEG encoder architecture processes word-level neural responses captured during stimulus presentation (see \ref{datasets_materials}). EEG input tensors $\mathbf{X} \in \mathbb{R}^{B \times L \times T \times C}$ represent batch size $B$, sequence length $L$ (number of words), temporal samples $T$, and channels $C$. Each word-level segment $\mathbf{x}_i \in \mathbb{R}^{T \times C}$ corresponds to neural activity recorded during processing of word $i$ in the stimulus sequence. For our primary experiments, we employ a transformer-based EEG encoder with 1 layer, 4 attention heads, and 768-dimensional hidden states. The encoder first reshapes word-level EEG segments into flat feature vectors:
\begin{equation}
\mathbf{x}_i^{\text{flat}} = \text{reshape}(\mathbf{x}_i) \in \mathbb{R}^{T \times C \rightarrow D}
\end{equation}
where $D = T \times C$ represents the flattened feature dimension. The sequence of word-level features $\mathbf{X}^{\text{flat}} = [\mathbf{x}_1^{\text{flat}}, \mathbf{x}_2^{\text{flat}}, \ldots, \mathbf{x}_L^{\text{flat}}]$ undergoes initial linear projection:
\begin{equation}
\mathbf{H}_0 = \mathbf{X}^{\text{flat}}\mathbf{W}_E + \mathbf{b}_E
\end{equation}
where $\mathbf{W}_E \in \mathbb{R}^{D \times d}$ and $\mathbf{b}_E \in \mathbb{R}^d$ project to hidden dimension $d=768$. The projected word sequence passes through transformer layers with self-attention mechanisms:
\begin{equation}
\mathbf{H}_l = \text{TransformerLayer}_l(\mathbf{H}_{l-1}), \quad l \in [1,2]
\end{equation}
yielding contextualised word-level representations $\mathbf{H}_2 \in \mathbb{R}^{L \times d}$ that capture both local neural patterns and sequential dependencies across the EEG word sequence. For sequence-level representation aggregation, we investigate four distinct pooling strategies to derive query and passage representations:

For sequence-level representation aggregation, we investigate four pooling strategies commonly employed in dense passage retrieval~\cite{karpukhin2020dense,lin2020distilling,lin2023aggretriever}. These methods represent the primary approaches for aggregating token-level representations into fixed-dimensional embeddings for query-passage matching:

\begin{enumerate}
\item \textbf{Max Pooling} computes element-wise maximum values across valid word positions, extracting the most salient neural features~\cite{lin2020distilling,lin2023aggretriever}:
\begin{equation}
\mathbf{h}_{\text{max}} = \max_{i=1}^{L} \mathbf{H}_2[i, :]
\end{equation}

\item \textbf{Mean Pooling} averages representations across valid positions for balanced semantic encoding~\cite{lin2020distilling}:
\begin{equation}
\mathbf{h}_{\text{mean}} = \frac{1}{L} \sum_{i=1}^{L} \mathbf{H}_2[i, :]
\end{equation}

\item \textbf{CLS Pooling} utilises a learnable classification token $\mathbf{h}_{\text{cls}}$ prepended to the sequence, following BERT-style aggregation~\cite{devlin2018bert} for unified sequence representation.

\item \textbf{Multi-vector} preserves word-level granularity by maintaining all contextualised representations $\{\mathbf{H}_2[i, :]\}_{i=1}^{L}$ for fine-grained similarity computation, following ColBERT-style dense retrieval~\cite{khattab2020colbert}.
\end{enumerate}

These four representation strategies enable the comparison of different semantic aggregation approaches for both EEG queries and text passages. Final representations undergo L2 normalisation before similarity computation in the shared embedding space, allowing direct comparison between neural and textual semantic content whilst preserving the distinct characteristics of each pooling strategy. We train each of the models using a contrastive learning objective adapted from InfoNCE~\cite{oord2018representation} for cross-sensory alignment. For each training batch containing $B$ query-passage pairs, we construct positive associations between corresponding EEG queries $\mathbf{q}_i$ and text passages $p_i$, whilst treating all other passages in the batch as negative examples. The contrastive loss encourages high similarity between positive pairs whilst pushing apart negative combinations:

\begin{equation}
\mathcal{L}_{\text{contrastive}} = -\frac{1}{B} \sum_{i=1}^{B} \log \frac{\exp(\text{sim}(\mathbf{q}_i, p_i) / \tau)}{\sum_{j=1}^{B} \exp(\text{sim}(\mathbf{q}_i, p_j) / \tau)}
\end{equation}

We employ InfoNCE~\cite{oord2018representation} for its proven effectiveness in learning cross-modal alignments by maximising agreement between positive pairs whilst minimising similarity to negative examples, making it particularly suitable for mapping between the disparate modalities of neural signals and text~\cite{karpukhin2020dense}. The similarity computation $\text{sim}(\mathbf{q}_i, p_j)$ depends on the pooling strategy:

Training employs the AdamW optimiser~\cite{loshchilov2017decoupled} with learning rate $1 \times 10^{-5}$, weight decay $0.01$, temperature parameter $\tau = 0.07$ and in-batch negatives with batchsize of 32. We implement linear warmup over 10 epochs followed by linear decay, with gradient clipping at maximum norm 1.0 for training stability. Early stopping with patience 10 epochs monitors validation loss to prevent overfitting. Models are trained for up to 200 epochs on NVIDIA A100 GPUs with mixed precision to accelerate training whilst maintaining numerical stability.

\section{Experimental Setup}
\label{experimental_setup}

Our experimental investigation addresses three core research questions regarding the effectiveness of EEG signals for B across sensory modalities:

\begin{itemize}
    \item \textbf{RQ1:} Can auditory EEG signals serve as effective passage retrieval queries, and how does their performance compare to visual EEG modalities in establishing viable brain-machine interface foundations?
    
    \item \textbf{RQ2:} Does combined training across auditory and visual EEG datasets improve retrieval performance, and are these benefits dependent on architectural choices for semantic aggregation?
    
    \item \textbf{RQ3:} Can combining EEG datasets collected under different modalities and protocols help address the limited availability of BPR training data, and do the benefits depend on architectural design choices?
\end{itemize}

\subsection{Evaluation Methodology}
\label{experimental_design}

To address these research questions, we implement an experimental design evaluating individual and combined modality training whilst controlling for confounding factors. Our approach comprises: (1) individual BPR models trained separately on auditory and visual EEG datasets, (2) a combined BPR model trained on merged datasets, and (3) evaluation across multiple architectural configurations. Each dataset (auditory Alice and visual Nieuwland) undergoes identical preprocessing with 80\% training, 10\% development, and 10\% test splits, maintaining consistent query-document distributions through dataset balancing procedures~\cite{mcguire2025towards}. Individual modality models employ identical hyperparameters and architectures as detailed in Section~\ref{model_architecture}, differing only in input EEG modality.

Our design incorporates four pooling strategies (max, mean, CLS, multi-vector) to evaluate semantic aggregation approaches across modalities~\cite{khattab2020colbert,devlin2018bert}. Individual models are evaluated on their respective test corpora, whilst the combined model is assessed on both unified and individual modality test sets, enabling direct comparison of combined versus individual training benefits (RQ2). To establish performance upper bounds, we include text-only baselines: BM25\footnote{BM25Okapi implementation from rank\_bm25 Python package} and ColBERTv2\footnote{ColBERTv2.0 from \texttt{colbert-ir/colbertv2.0}}~\cite{khattab2020colbert}, which retrieve using actual text queries rather than EEG.

To assess model robustness, we evaluate across document masking ratios (0\%, 25\%, 50\%, 75\%, 90\%, 100\%) to simulate varying information scarcity~\cite{karpukhin2020dense,lee2019latent}, revealing whether models rely on lexical overlap or capture deeper semantic relationships. We report performance using IR metrics: \textbf{Hit@k} (k $\in$ \{1, 5, 10\}), indicating whether the relevant document appears within top-k results~\cite{karpukhin2020dense}, and \textbf{Mean Reciprocal Rank (MRR)}, computed as $\text{MRR} = \frac{1}{|Q|} \sum_{i=1}^{|Q|} \frac{1}{\text{rank}_i}$, where $\text{rank}_i$ is the relevant document's rank for query $i$~\cite{robertson2009probabilistic}.

\section{Results \& Discussion}
\begin{table*}[h]
\centering
\caption{Performance comparison across pooling methods. Values show mean ± std across masking levels (0\%, 25\%, 50\%, 75\%, 90\%, 100\%). \textcolor{gray}{Text baselines (greyed) represent upper bound.} Best neural results per column in bold. \underline{Underlined} indicates combined training significantly outperforms individual training (paired t-test, p < 0.05).}
\label{tab:pooling_comparison}
\footnotesize
\setlength{\tabcolsep}{3.5pt}
\renewcommand{\arraystretch}{0.85}
\begin{tabular}{lllcccc}
\toprule
\textbf{Training} & \textbf{Pooling} & \textbf{Eval Data} & \textbf{MRR} & \textbf{Hit@1} & \textbf{Hit@5} & \textbf{Hit@10} \\
\midrule
\multicolumn{7}{l}{\textit{\textcolor{gray}{Text-Only Baselines (Upper Bound)}}} \\
\midrule
\multirow{2}{*}{\textcolor{gray}{\textbf{—}}} & \multirow{2}{*}{\textcolor{gray}{\textbf{BM25}}} & \textcolor{gray}{Audio} & \textcolor{gray}{.428{\scriptsize±.338}} & \textcolor{gray}{.342{\scriptsize±.365}} & \textcolor{gray}{.499{\scriptsize±.337}} & \textcolor{gray}{.542{\scriptsize±.301}} \\
 & & \textcolor{gray}{Visual} & \textcolor{gray}{.420{\scriptsize±.350}} & \textcolor{gray}{.339{\scriptsize±.371}} & \textcolor{gray}{.482{\scriptsize±.349}} & \textcolor{gray}{.523{\scriptsize±.315}} \\
\cmidrule{2-7}
 & \multirow{2}{*}{\textcolor{gray}{\textbf{ColBERT}}} & \textcolor{gray}{Audio} & \textcolor{gray}{.296{\scriptsize±.189}} & \textcolor{gray}{.178{\scriptsize±.176}} & \textcolor{gray}{.407{\scriptsize±.216}} & \textcolor{gray}{.526{\scriptsize±.198}} \\
 & & \textcolor{gray}{Visual} & \textcolor{gray}{.490{\scriptsize±.210}} & \textcolor{gray}{.319{\scriptsize±.263}} & \textcolor{gray}{.710{\scriptsize±.156}} & \textcolor{gray}{.809{\scriptsize±.109}} \\
\midrule
\multicolumn{7}{l}{\textit{Neural EEG-to-Text Retrieval Models}} \\
\midrule
\textbf{Random} & \textbf{NOISE} & — & .010{\scriptsize±.002} & .002{\scriptsize±.001} & .006{\scriptsize±.003} & .013{\scriptsize±.005} \\
\midrule
\multirow{8}{*}{\textbf{Individual}} & \multirow{2}{*}{\textbf{MEAN}} & Audio & .162{\scriptsize±.048} & .075{\scriptsize±.028} & .233{\scriptsize±.071} & .340{\scriptsize±.102} \\
 & & Visual & .073{\scriptsize±.032} & .033{\scriptsize±.015} & .092{\scriptsize±.042} & .137{\scriptsize±.068} \\
\cmidrule{2-7}
 & \multirow{2}{*}{\textbf{MAX}} & Audio & .348{\scriptsize±.038} & .212{\scriptsize±.027} & .494{\scriptsize±.058} & .634{\scriptsize±.069} \\
 & & Visual & .198{\scriptsize±.061} & .119{\scriptsize±.054} & .271{\scriptsize±.065} & .338{\scriptsize±.084} \\
\cmidrule{2-7}
 & \multirow{2}{*}{\textbf{MULTI}} & Audio & .234{\scriptsize±.064} & .131{\scriptsize±.044} & .329{\scriptsize±.093} & .446{\scriptsize±.119} \\
 & & Visual & .180{\scriptsize±.059} & .111{\scriptsize±.051} & .249{\scriptsize±.060} & .300{\scriptsize±.069} \\
\cmidrule{2-7}
 & \multirow{2}{*}{\textbf{CLS}} & Audio & .362{\scriptsize±.062} & .220{\scriptsize±.051} & .523{\scriptsize±.088} & .668{\scriptsize±.092} \\
 & & Visual & .139{\scriptsize±.052} & .074{\scriptsize±.031} & .192{\scriptsize±.071} & .262{\scriptsize±.098} \\
\midrule
\multirow{8}{*}{\textbf{Combined}} & \multirow{2}{*}{\textbf{MEAN}} & Audio & .165{\scriptsize±.007} & \underline{.092}{\scriptsize±.007} & .212{\scriptsize±.008} & .291{\scriptsize±.007} \\
 & & Visual & \underline{.264}{\scriptsize±.022} & \underline{.151}{\scriptsize±.016} & \underline{.353}{\scriptsize±.029} & \underline{.523}{\scriptsize±.027} \\
\cmidrule{2-7}
 & \multirow{2}{*}{\textbf{MAX}} & Audio & .268{\scriptsize±.009} & .180{\scriptsize±.014} & .335{\scriptsize±.003} & .443{\scriptsize±.003} \\
 & & Visual & \underline{\textbf{.465}}{\scriptsize±.041} & \underline{\textbf{.313}}{\scriptsize±.026} & \underline{\textbf{.646}}{\scriptsize±.066} & \underline{\textbf{.803}}{\scriptsize±.058} \\
\cmidrule{2-7}
 & \multirow{2}{*}{\textbf{MULTI}} & Audio & .231{\scriptsize±.010} & \underline{.153}{\scriptsize±.012} & .289{\scriptsize±.004} & .378{\scriptsize±.009} \\
 & & Visual & \underline{.319}{\scriptsize±.008} & \underline{.190}{\scriptsize±.008} & \underline{.440}{\scriptsize±.017} & \underline{.601}{\scriptsize±.007} \\
\cmidrule{2-7}
 & \multirow{2}{*}{\textbf{CLS}} & Audio & \textbf{\underline{.474}}{\scriptsize±.020} & \textbf{\underline{.314}}{\scriptsize±.017} & \textbf{\underline{.671}}{\scriptsize±.046} & \textbf{\underline{.858}}{\scriptsize±.034} \\
 & & Visual & \underline{.256}{\scriptsize±.014} & \underline{.141}{\scriptsize±.019} & \underline{.351}{\scriptsize±.005} & \underline{.515}{\scriptsize±.008} \\
\bottomrule
\end{tabular}
\end{table*}

\begin{figure}
    \centering
    \includegraphics[width=0.9\linewidth]{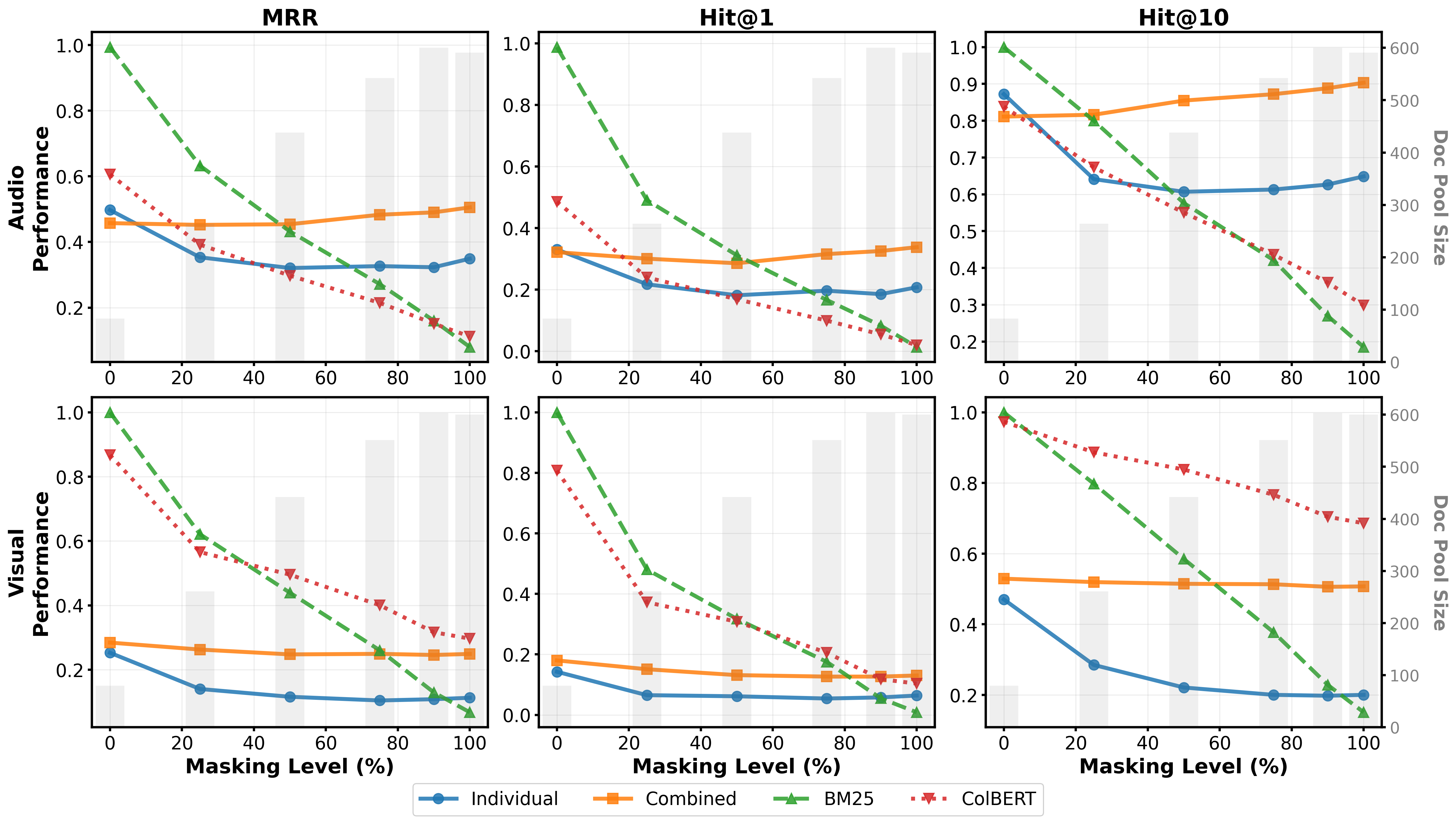}
    \caption{cross-sensory training effects comparing individual versus combined training for Auditory (Alice, top row) and Visual(Nieuwland, bottom row) datasets across MRR, Hit@1, and Hit@10 metrics. Individual training (blue) uses only modality-specific data, whilst combined training (orange) leverages both auditory and visual datasets. Grey bars show document pool sizes at each masking level.}
    \label{fig:modality_masking}
\end{figure}

\subsection{Auditory EEG Demonstrates Effective Brain Passage Retrieval Performance}

Addressing RQ1 regarding the viability of auditory EEG signals for BPR, our results establish that auditory modalities achieve substantial retrieval effectiveness. As shown in Table~\ref{tab:pooling_comparison}, auditory EEG models achieve strong performance across all architectural configurations, with CLS pooling reaching MRR of 0.362, Hit@1 of 0.220, and Hit@10 of 0.668 in individual training. MAX pooling similarly demonstrates effectiveness with MRR of 0.348, Hit@1 of 0.212, and Hit@10 of 0.634. These results represent 36-fold improvements over random noise baselines (MRR: 0.010), confirming that meaningful semantic information persists in auditory neural signals rather than spurious statistical patterns~\cite{jo2024eeg}.

Comparative analysis reveals that auditory signals achieve 160\% higher MRR, 197\% higher Hit@1, and 155\% higher Hit@10 than visual signals under CLS pooling, with consistent advantages across other architectures. Figure~\ref{fig:modality_masking} demonstrates that auditory EEG maintains stable performance across masking levels, with combined training showing particular robustness, whilst text baselines degrade substantially at high masking ratios (75-100\%), auditory EEG sustains retrieval effectiveness, indicating semantic encoding rather than reliance on lexical overlap. Notably, combined training with CLS pooling enables auditory EEG to exceed BM25 (MRR: 0.474 vs 0.428; Hit@5: 0.671 vs 0.499; Hit@10: 0.858 vs 0.542) and ColBERT across all metrics. Whilst ColBERT was not fine-tuned on our specific ICT task formulation, this suggests that EEG-based retrieval may be capable of competing with strong lexical/neural text baselines. Figure~\ref{fig:modality_masking} reveals that this advantage becomes most pronounced at high masking levels, where EEG-based retrieval maintains effectiveness (Hit@10: ~0.65 at 75\% masking) whilst BM25 and ColBERT degrade to below 0.4.

These findings establish auditory EEG as a viable modality for accessible IR systems, representing the first demonstration of EEG queries competing with and in some cases exceeding text-based retrieval. This capability could enable BMIs for users with visual impairments, who represent approximately 285 million people worldwide~\cite{who2019world} and face substantial barriers in traditional text-based search systems~\cite{bigham2017effects,moshfeghi2025brain}, as well as users with motor disabilities affecting traditional input methods. Additionally, auditory-based neural interfaces align naturally with emerging audio-centric retrieval scenarios such as conversational search systems, voice-activated interfaces, and retrieval over spoken content (podcasts, audiobooks) where visual stimuli are absent yet users form information needs~\cite{radlinski2017theoretical}. The relative robustness of EEG-based retrieval at high masking levels suggests potential for neural query interfaces that could reduce formulation barriers whilst maintaining competitive retrieval effectiveness, though further investigation with task-specific fine-tuned text baselines would be valuable to fully characterise this capability.

\subsection{Cross-Sensory Training Benefits Manifest Through Architecture-Dependent Mechanisms}

\begin{figure}
    \centering
    \includegraphics[width=\linewidth]{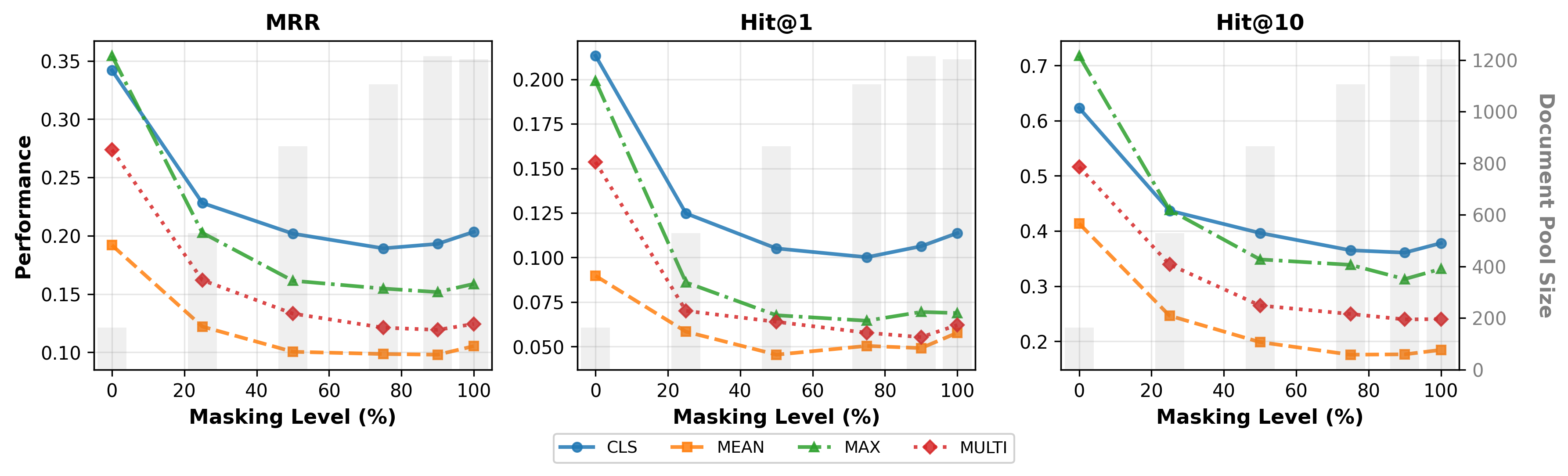}
    \caption{Architectural comparison of pooling strategies (CLS, MEAN, MAX, MULTI) on combined training data across three key metrics (MRR, Hit@1, Hit@10). Performance is evaluated at six masking levels (0-100\%). Grey bars indicate document pool sizes.}
    \label{fig:pooling_masking}
\end{figure}

Addressing RQ2, combined training effects vary across pooling architectures. CLS pooling demonstrates bidirectional improvements: auditory performance increases by 31\% in MRR (0.362$\rightarrow$0.474), 43\% in Hit@1 (0.220$\rightarrow$0.314), and 28\% in Hit@10 (0.668$\rightarrow$0.858), whilst visual performance improves by 84\% in MRR (0.139$\rightarrow$0.256), 91\% in Hit@1 (0.074$\rightarrow$0.141), and 97\% in Hit@10 (0.262$\rightarrow$0.515). Figure~\ref{fig:pooling_masking} demonstrates that CLS pooling maintains robust performance across all masking levels for both modalities, with this bidirectional enhancement yielding the highest overall BPR performance (MRR: 0.474, Hit@10: 0.858).

Alternative pooling strategies exhibit asymmetric benefits. MAX pooling improves visual performance substantially (135\% MRR increase: 0.198$\rightarrow$0.465; 163\% Hit@1 increase: 0.119$\rightarrow$0.313; 137\% Hit@10 increase: 0.338$\rightarrow$0.803) whilst reducing auditory performance by 23\% in MRR. MULTI pooling shows similar asymmetry with 77\% visual MRR improvement but minimal auditory changes. Figure~\ref{fig:pooling_masking} reveals these architectural differences across masking conditions, suggesting that learned sequence-level aggregation through CLS tokens more effectively leverages cross-modal information than statistical pooling approaches~\cite{devlin2018bert}.

These findings directly address RQ3 by revealing that the benefits of combining diverse EEG datasets depend critically on architectural design choices. As evidenced by the results above, different pooling strategies prove optimal for different sensory modalities: CLS pooling achieves bidirectional improvements across both MRR and Hit metrics, whilst MAX pooling substantially benefits visual performance but reduces auditory effectiveness. This architecture-modality interaction suggests that optimal BPR systems should employ stimulus-specific pooling strategies rather than universal designs. The success of combined training despite minimal lexical overlap between datasets (Jaccard similarity: 0.175-0.184) demonstrates that diverse EEG corpora can address data scarcity challenges~\cite{hollenstein2018zuco,bhattasali2020alice}, but practitioners must carefully match aggregation mechanisms to the target sensory modality and deployment context to realise these benefits.

\section{Conclusion}
\label{conclusion}
We presented the first investigation of auditory EEG signals for brain passage retrieval. Addressing RQ1, our findings demonstrate that auditory EEG serves as effective passage retrieval queries, with combined training achieving performance that competes with and in some cases exceeds traditional text baselines. This establishes auditory neural patterns as viable query representations for BMIs with implications for accessible IR, particularly for users with visual impairments and motor disabilities, as well as audio-centric contexts where visual stimuli are absent, such as podcast search, audiobook retrieval, and conversational search interfaces. Addressing RQ2 and RQ3, we demonstrated that combined training improves performance through architecture-dependent mechanisms, with optimal pooling strategies differing by modality, helping address EEG dataset scarcity by showing that combining diverse corpora benefits both modalities despite minimal lexical overlap. Our investigation has important limitations. We utilised EEG from naturalistic reading and listening rather than active information seeking, and evaluated across only two datasets with different source texts, limiting our ability to isolate cross-sensory neural contributions from general dataset diversity effects. Future work should investigate whether these findings generalise to EEG captured during actual query formulation and information need realisation, and expand the evaluation to additional datasets. 

\section*{Disclosure of Interests}
The authors have no competing interests to declare that are relevant to the content of this article.

\begin{acks}
This work was supported by the Engineering and Physical Sciences Research Council [grant number EP/W522260/1]. Results were obtained using the ARCHIE-WeSt High Performance Computer\footnote{www.archie-west.ac.uk} based at the University of Strathclyde.
\end{acks}

\bibliographystyle{splncs04}
\bibliography{ref}  % Replace with your .bib filename (without .bib extension)

\end{document}